\documentclass[useAMS,usenatbib]{mn2e}

\usepackage{epsfig,color}
\usepackage{natbib}
\bibpunct{(}{)}{;}{a}{}{,}
\def\spose#1{\hbox to 0pt{#1\hss}}
\def\ltsimm{\mathrel{\spose{\lower 3pt\hbox{$\sim$}}
        \raise 2.0pt\hbox{$<$}}}
\def\gtsimm{\mathrel{\spose{\lower 3pt\hbox{$\sim$}}
        \raise 2.0pt\hbox{$>$}}}

\def\km{{\rm\thinspace km}}
\def\cm{{\rm\thinspace cm}}
\def\pc{{\rm\thinspace pc}}

\def\s{{\rm\thinspace s}}

\def\g{{\rm\thinspace g}}
\def\kmps{\hbox{${\rm\km\s^{-1}\,}$}}

\def\erg{{\rm\thinspace erg}}

\def\Hz{{\rm\thinspace Hz}}

\def\ster{{\rm\thinspace ster}}
\def\ergps{\hbox{${\rm\erg\s^{-1}\,}$}}

\def\Msol{\hbox{${\rm\thinspace M_{\odot}}$}}

\def\pcm{\hbox{${\rm\cm^{-1}\,}$}}
\def\pcm2{\hbox{${\rm\cm^{-2}\,}$}}
\def\pcm3{\hbox{${\rm\cm^{-3}\,}$}}
\def\ergpscm3Hz{\hbox{${\rm\ergps\cm^{-3}\Hz^{-1}\,}$}}
\def\ergpscm3Hzster{\hbox{${\rm\ergps\cm^{-3}\Hz^{-1}\ster^{-1}\,}$}}
\def\gpcm3{\hbox{${\rm\g\cm^{-3}\,}$}}
\def\ergpcm2{\hbox{${\rm\erg\cm^{-2}\,}$}}
\def\ergpcm3{\hbox{${\rm\erg\cm^{-3}\,}$}}
\def\phpscm2{\hbox{${\rm photons\s^{-1}\cm^{-2}\,}$}}

\title[Tails of the Unexpected]{Tails of the Unexpected: The Interaction of an Isothermal Shell with a Cloud}
\author[J.~M.~Pittard]
{J. M. Pittard$^{1}$\thanks{E-mail: jmp@ast.leeds.ac.uk}\\
$^{1}$School of Physics and Astronomy, The University of
        Leeds, Woodhouse Lane, Leeds LS2 9JT, UK
}

\begin{document}

\date{Accepted ... Received ...; in original form ...}

\pagerange{\pageref{firstpage}--\pageref{lastpage}} \pubyear{2011}

\maketitle

\label{firstpage}

\begin{abstract}
A new mechanism for the formation of cometary tails behind dense
clouds or globules is discussed. Numerical hydrodynamical models show
that when a dense shell of swept-up matter overruns a cloud, material
in the shell is focussed behind the cloud to form a tail. This mode of
tail formation is completely distinct from other methods, which
involve either the removal of material from the cloud, or shadowing
from a strong, nearby source of ionization. This mechanism is relevant
to the cometary tails seen in planetary nebulae and to the
interaction of superbubble shells with dense clouds.
\end{abstract}


\section{Introduction}
\label{sec:intro}
An extensive literature of analytical and numerical investigations of
shock-cloud interactions now exists in which the effects of magnetic
fields, radiative cooling, thermal conduction and turbulence have all
been considered \citep*[see, e.g.,][and references
  therein]{Pittard:2009,Pittard:2010}. However, in many astrophysical
objects the cloud size is comparable to or larger than the depth of
the post-shock layer, so it is surprising that to date almost all
numerical investigations have been performed in the small-cloud limit
where the post-shock flow has effectively an infinite depth. Notable
exceptions include investigations where the global flow is simulated,
for instance, in the interaction of a supernova remnant (SNR) with a
cloud \citep{Tenorio-Tagle:1986,Rozyczka:1987,Leao:2009}, though such
models suffer from low numerical resolution.  Higher resolution was
used in work on the triggered collapse of a cloud by a thin shell
\citep{Boss:2010}, but the downstream flow was not examined.
In this paper we re-examine the interaction of a cold, dense,
isothermal shell with a spherical cloud, studying the effect of the
shell thickness and Mach number. 

\section{Shell-cloud Interaction} 
The shell is driven by a high pressure, low density, hot
bubble. Pressure-driven shells occur in planetary nebulae (PNe) and
around individual stars or groups of massive stars, and have diameters
ranging from $\sim 0.1\;$pc to a few kpc. Since cooling breaks the
scale-free nature of adiabatic simulations one is forced to choose a
particular lengthscale. In this paper the interaction of a superbubble
with a molecular cloud is simulated, though the results are also
relevant to the smaller-scale interactions mentioned above.

The structure of the superbubble, including the thickness of the shell
and the density contrast between the shell and the hot interior, can
be obtained from a global hydrodynamical simulation. However, the shell
properties depend on assumptions about the structure of the ISM (e.g.,
whether there is a density gradient, the strength and orientation of
magnetic fields, the number, size and density distributions of clouds,
etc.)  and the efficiency of mass transfer from the cool shell and
from clouds within the hot bubble to the hot, rarefied gas. Therefore,
we choose instead to specify instantaneous shell and bubble properties
(Mach number, surface density, bubble temperature) in order to focus
on the key physics of the interaction.  The shell is assumed to be
planar (i.e. its radius is much larger than the cloud).  

Our investigation is based around a standard model with the following
parameters. The inter-cloud ISM has an assumed density and temperature
of $\rho_{\rm amb} = 3.33 \times 10^{-25}\;\gpcm3$ and $T_{\rm
  amb}=8000\;{\rm K}$. The density contrast of the spherical cloud
$\chi = 10^{3}$, and its core temperature is $T_{\rm c} = 8\;{\rm
  K}$. The cloud is thus in pressure equilibrium with its
surroundings. To keep matters as simple as possible in this first
investigation all material is assumed to behave as an ideal gas with a
ratio of specific heats, $\gamma=5/3$, although in reality the cores
are molecular. This simplification has little effect on the overall
dynamics.  The ISM pressure $p_{\rm amb}=2667 \;\pcm3{\rm K}$.  The
cloud has a mass $M_{\rm c}=300 \Msol$ and a radius $r_{\rm c}=2\pc$.
The density within the hot bubble $\rho_{\rm bub}=\rho_{\rm amb}/100$
ensures that this material does not appreciably cool during the
simulations. We ignore thermal conduction, magnetic fields and
self-gravity. Though the latter is important for the parameters we
have chosen, our focus in this paper is on the development of a tail
behind the cloud, rather than star formation within it.

\begin{figure}
\begin{center}
\psfig{figure=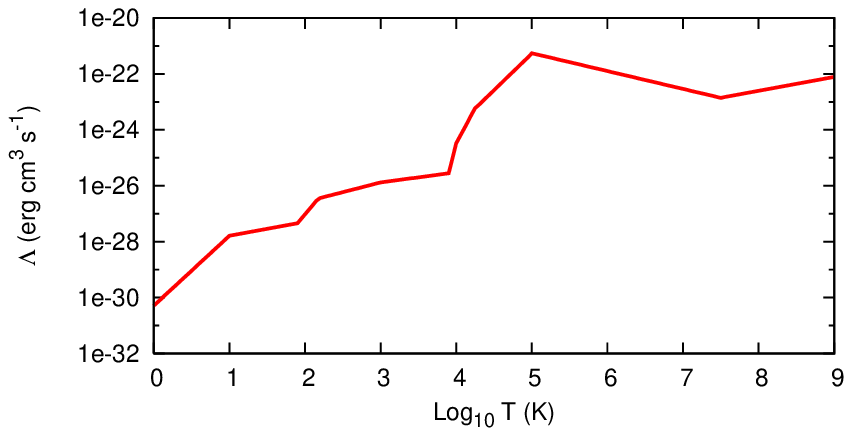,width=7.5cm}
\psfig{figure=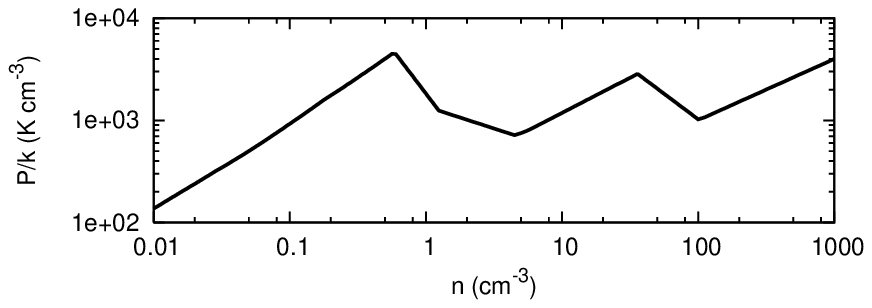,width=7.5cm}
\end{center}
\caption[]{Top: The parameterized cooling curve adopted in this work.
Bottom: The density-pressure phase diagram corresponding to the cooling
curve shown and $\Gamma = 10^{-26}\ergps$.}
\label{fig:cooling}
\end{figure}

The net heating/cooling rate per unit volume is parameterized as
$\dot{e} = n\Gamma - n^{2}\Lambda$, where $n=\rho/m_{\rm
H}=1.43\,n_{\rm H}$, and $\Gamma$ and $\Lambda$ are heating and
cooling coefficients which are assumed to depend only on temperature.
In the ISM, $\Gamma$ decreases with increasing density as the
starlight, soft X-ray, and cosmic ray flux are attenuated by the high
column densities associated with dense clouds. Because the exact form
of the attenuation depends on details which remain uncertain (for
instance the size and abundance of PAHs), the heating rate at $T
\ltsimm 10^{4}\;{\rm K}$ is similarly uncertain. In this work we
assume that $\Gamma = 10^{-26}\ergps$ (independent of $\rho$ or $T$).
The low-temperature ($T \ltsimm 10^{4}\;{\rm K}$) cooling was
then adjusted to give 3 thermally stable phases at thermal pressures between
$2000-6000 \;\pcm3{\rm K}$, as required by observations. These stable
phases, at temperatures $\sim 10\,$K, $\sim 150\,$K and $\sim
8500\,$K, correspond to the molecular, atomic and warm
neutral/ionized phases, respectively. The cooling curve and phase
diagram are shown in Fig.~\ref{fig:cooling}. The heating and cooling
rates at low temperatures are higher than in reality, but since low
temperature material is usually at high density (to maintain pressure
balance) the timescale to reach equilibrium is short, and the simulations
presented are not expected to be significantly affected by these
simplifications.

The pressure of the shell is high enough that the gas within it should
be in the cold atomic or molecular phase. However, in all the
simulations the column density of the shell is much less than that of
the molecular cloud, so it is assumed that the heating is high enough
to keep the shell material in the warm ionized phase.  Therefore, the
temperature of material in the shell is prevented from cooling below
$8000\;{\rm K}$, and the shock sweeping through the ISM behaves
isothermally. In contrast, material ablated from the cloud is allowed
to find its own equilibrium temperature, which may be higher or lower
than $T_{\rm amb}$.

We adopt the density profile noted in \citet{Pittard:2009} with
$p_{1}=10$ (i.e. a cloud with a reasonably sharp edge).  Although the
cloud is initially in pressure equilibrium with its surroundings, the
adopted cooling curve means that the edge of the cloud is not in a
stable thermal configuration. However, the shell hits the cloud before
it has time to substantially adjust itself, and the initial transient
is unimportant to the following results. In contrast, the intercloud
ISM is close to thermal equilibrium and its temperature remains almost
constant throughout the simulation.

A key parameter for the destruction of non-magnetic, non-conducting,
non-gravitating adiabatic clouds is the cloud-crushing timescale
\citep*{Klein:1994},
\begin{equation}
\label{eq:tcc}
t_{\rm cc} \equiv \frac{\chi^{1/2}r_{\rm c}}{v_{\rm s}} = \frac{\chi^{1/2}r_{\rm c}}{M_{\rm a}a},
\end{equation}
where $v_{\rm s}$ and $M_{\rm a}$ are the speed and adiabatic Mach
number of the shock through the intercloud medium having an adiabatic
sound speed $a$. Although
our study involves radiative shocks, the above characteristic
timescale remains useful and is used in our analysis. All times are
measured relative to when the shell is level with the cloud centre.

\begin{figure*}
\begin{center}
\psfig{figure=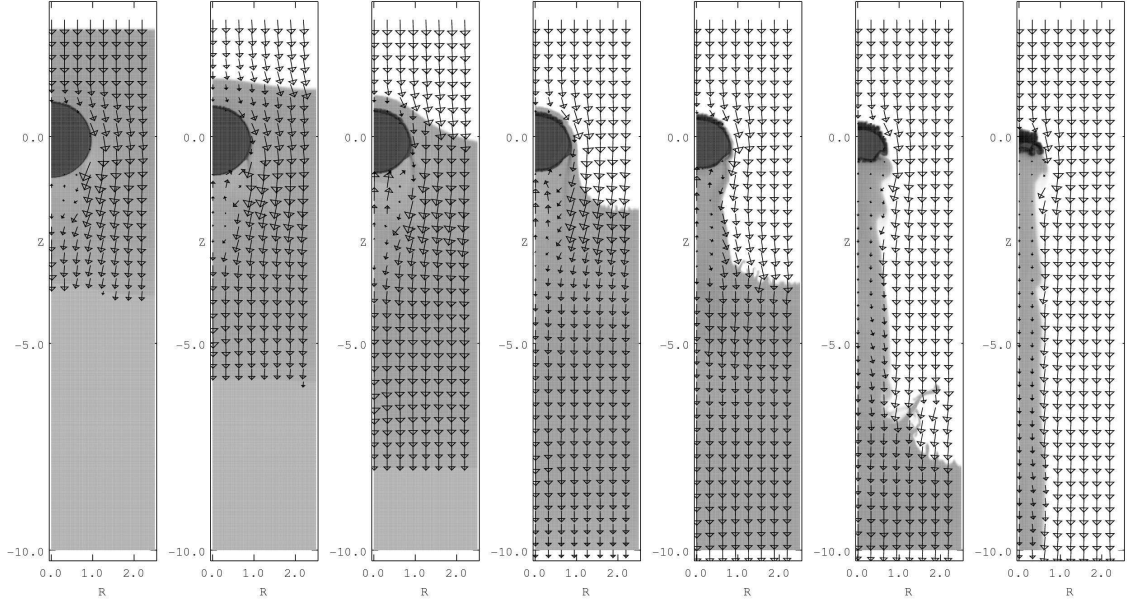,width=15.0cm}
\end{center}
\caption[]{The time evolution of a 2D $rz$-axisymmetric shell-cloud
interaction with $M=1.5$, $\chi=10^{3}$ and an initial shell
thickness of $8\,$pc ($4\,r_{c}$). The interaction proceeds left to
right with $t=0.098$, 0.162, 0.227, 0.291, 0.355, 0.549 and
$0.806\,t_{\rm cc}$.  Material in the shell is focussed into a tail as
it passes over the cloud which is compressed due to the jump in pressure.  
The density maps span over 5 dex: the bubble interior has $n_{\rm H} =
1.5\times10^{-3}\,\pcm3$ (white), the ambient medium, shell and
tail have $n_{\rm H} = 0.15-0.6\,\pcm3$ (light grey), and the
unshocked cloud has $n_{\rm H} = 150\,\pcm3$ (dark grey). The highest speed
attained by the flow is $23.8\,\kmps$. The unit of length is 2\,pc.}
\label{fig:rz12}
\end{figure*}

\begin{figure*}
\begin{center}
\psfig{figure=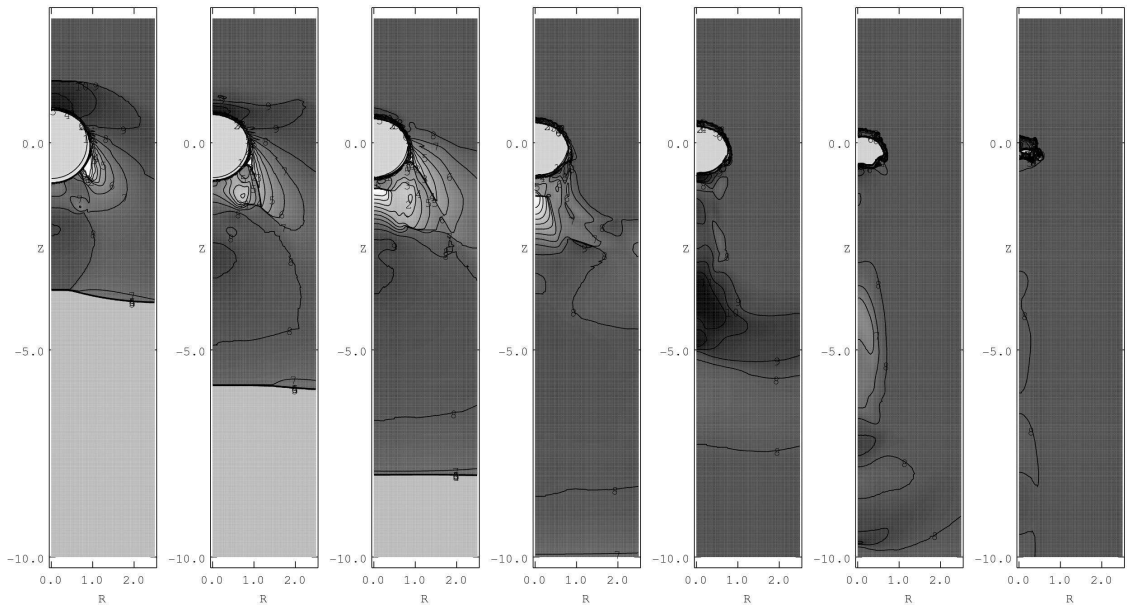,width=15.0cm}
\end{center}
\caption[]{As Fig.~\ref{fig:rz12} but showing the log of the pressure (black is high) with contour levels every 0.1\,dex. The pressure gradients which exist around the cloud in the initial stages of the interaction focus material from the shell onto the symmetry axis.}
\label{fig:rzpre}
\end{figure*}

\section{Numerical Simulations}
\label{sec:results}
We use the MG code recently developed by Falle. It employs an exact
Riemann solver for gasdynamics and piece-wise linear interpolation to
achieve second order accuracy in space and time
\citep[cf.][]{Falle:1991}.  Adapative mesh refinement is handled on a
cell-by-cell basis, and it is fully parallelized using MPI.  Physics
modules, including ones to handle self-gravity and magnetic fields,
can be turned on or off as required.  One such module is a
$k$-$\epsilon$ subgrid model, which we use in this work to calculate
the properties of the turbulence which is generated in high Reynolds
number astrophysical flows. Such models simulate the effect that the
turbulent eddies have on the mean flow by increasing the transport
coefficients, particularly the viscosity, in regions where there is a
lot of turbulence. Further details including the full set of equations
solved can be found in \citet{Pittard:2009}.  The 2D calculations are
computed on an $rz$ axisymmetric grid, with a domain of $0 \leq r \leq
24$, $-60 \leq z \leq 100$ for all models. 8 grid levels are used with
128 cells per cloud radius on the finest grid. A 3D calculation with a
domain of $-30 \leq x \leq 30$, $0 \leq y \leq 3$, $0 \leq z \leq 3$
and 64 cells per cloud radius was also performed. The cloud is
initially centered at the origin.

\begin{figure}
\begin{center}
\psfig{figure=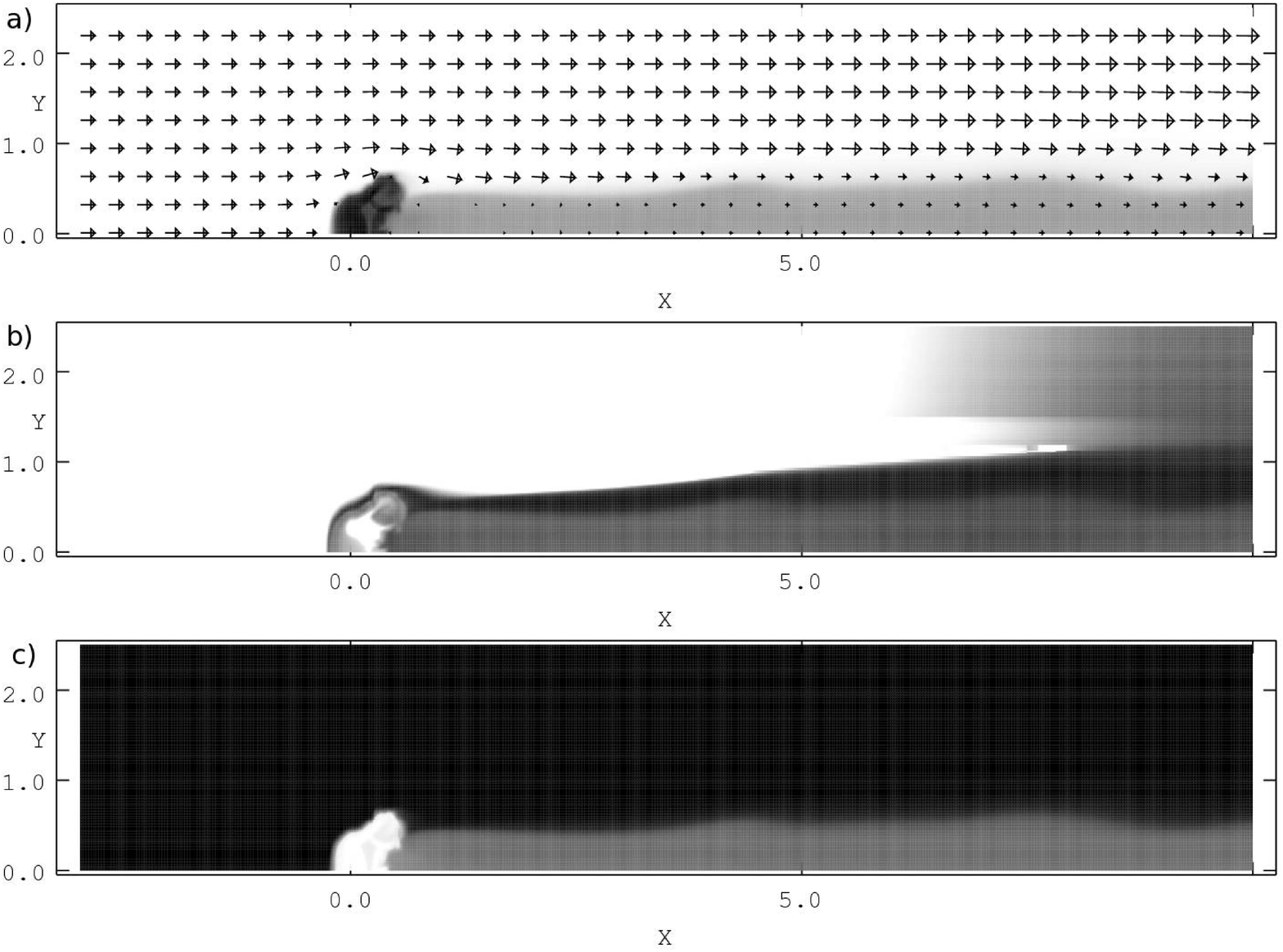,width=8.5cm}
\end{center}
\caption[]{Same as Fig.~\ref{fig:rz12}, but showing cuts through the
  $z=0$ plane of a 3D simulation at $t = 0.806\,t_{\rm cc}$. The shell
  is moving in the $+x$ direction, and there is reflection symmetry
  about the $y=0$ plane.  The panels show logarithmic values (black is
  high) of: a) the mass density and $xy$-velocity vectors; b) the
  turbulent energy density per unit mass; c) the temperature.}
\label{fig:3d}
\end{figure}

\begin{figure}
\begin{center}
\psfig{figure=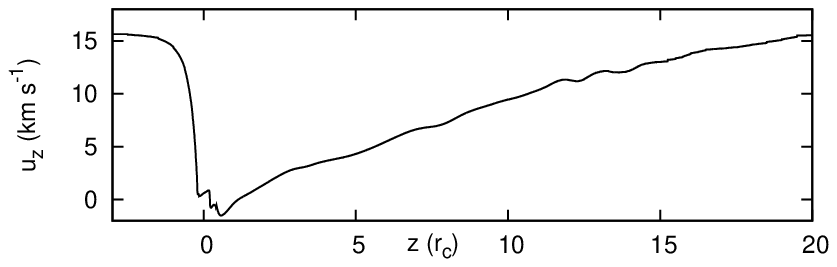,width=8.5cm}
\end{center}
\vspace{-5mm}
\caption[]{Velocity along the core of the tail at $t = 0.806\,t_{\rm
    cc}$ for the interaction shown in Fig.~\ref{fig:rz12}.  The cloud
  remains near $z=0.0$. The subsonic flow in the interior of the
  bubble (from the left in this plot) decelerates as it approaches the
  cloud.}
\label{fig:vel_profile}
\end{figure}

Fig.~\ref{fig:rz12} shows the time evolution of the interaction of a
2D model with an isothermal shell of thickness $l_{\rm sh}=8\,$pc and
Mach number $M_{\rm a} = 1.5$.  The pressure and density jump in the
shell is $\gamma M_{\rm a}^{2} = M_{\rm i}^{2} = 3.75$ (where $M_{\rm
  a}$ and $M_{\rm i}$ are respectively the adiabatic and isothermal
Mach numbers of the shell).  The ratio of the column density through
the shell and the centre of the cloud, $\sigma_{\rm sh}/\sigma_{\rm
  cl} = 7.5 \times 10^{-3}$. Hence the cloud is relatively unaffected
by the initial passage of the shell, though the shocks driven into it
by the jump in the external pressure start to compress it. This
compression is nearly isotropic, due to the low Mach number of the
interaction and the high density contrast of the cloud. The cloud then
exits through the back surface of the shell to reside in the low
density interior of the hot bubble driving the shell.

The most interesting aspect of the interaction is the formation of a
tail behind the cloud \citep[for earlier simulations which display
  similar but much thinner tails see][]{Tenorio-Tagle:1984b}. The tail
is mainly composed of material from the shell, with only small amounts
(less than a few percent concentration) of material ablated or
stripped from the cloud. The part of the shell adjacent to the cloud
moves in the lateral direction onto the axis due to the pressure
gradient which exists across its face as it sweeps over the cloud (see
Fig.~\ref{fig:rzpre}). A large eddy forms which causes this material
to lose speed relative to the rest of the shell, with some material
near the axis flowing back towards the rear of the cloud \citep*[see
  also][]{Nittmann:1982,Tenorio-Tagle:1984a}. This material is
subsequently compressed against the axis by the hot, subsonic flow
which overtakes it.  The pressure gradient across the face of the
shell diminishes as the shell moves further downstream, and the
focusing becomes more gradual. 
A 3D simulation of this interaction reveals the same
features (see Fig.~\ref{fig:3d}), indicating that this is a robust
result which is not dependent on the assumed axisymmetry. The high
shear around the cloud causes a turbulent boundary layer at the edge
of the tail which grows with an opening angle of
$\approx3-4^{\circ}$. The interior parts of the tail also contain some
turbulence, though the central part of the cloud has none.

The tails in these models exhibit a large length-to-width ratio which
can reach nearly 50:1 at late times ($t = 2.1\,t_{\rm cc}$).
Fig.~\ref{fig:vel_profile} shows the velocity profile along the
symmetry axis through the core of the tail at $t = 0.806\,t_{\rm cc}$
for the interaction shown in Fig.~\ref{fig:rz12}. 
The acceleration is approximately constant along the
tail, with the velocity reaching $\approx 10\,\kmps$ ($M_{\rm a} = 0.8$) at $z
= 10\,r_{\rm c}$. 
Due to the lack of material being stripped off the cloud, the tail
eventually dissipates as it thins and then detaches from the cloud (by
$t \sim 3\,t_{\rm cc}$). The axial velocity profile perpendicular to
the tail shows a near constant speed at a given downstream position,
which indicates efficient momentum transfer across the tail.

\begin{figure*}
\begin{center}
\psfig{figure=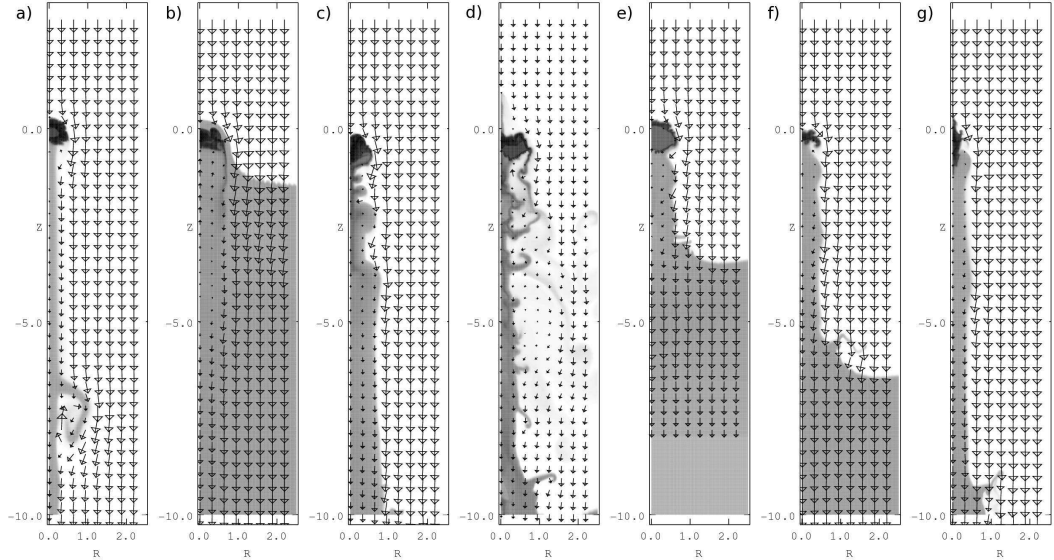,width=14.0cm}
\end{center}
\caption[]{(a) $M=1.5$,
$l_{\rm sh}=2\,$pc ($r_{\rm c}$), $t=0.806\,t_{\rm cc}$. (b) and (c): $M=1.5$, $l_{\rm sh}=32\,$pc, $t=0.806$ and
$1.192\,t_{\rm cc}$. (d) $M=3.0$, $l_{\rm sh}=8\,$pc, $t=0.519\,t_{\rm
cc}$. (e)$-$(g): $\chi=125$, $M=1.5$, $l_{\rm sh}=2\,$pc ($r_{\rm c}$),
$t = 0.642$, 1.006, and $1.552\,t_{\rm cc}$. Note that the cloud is
thermally unstable in this model, and collapses to higher densities
and much colder temperatures than its initial conditions. The 
density grey-scale is consistent in all panels, but the velocity vectors in
panel d) are twice as short for a given speed compared to the other panels.}
\label{fig:various}
\end{figure*}

We have performed a series of models designed to explore parameter
space to determine the conditions necessary for tail production (see
Fig.~\ref{fig:various}). Decreasing the thickness of the shell (i.e.
$\sigma_{\rm sh}/\sigma_{\rm cl}$) leads to a thinner tail
(Fig.~\ref{fig:various}a).  Increasing the thickness of the shell
enhances the stripping of material from the cloud, which causes
oscillations in the tail width (Figs.~\ref{fig:various}b and
c). Interactions at higher Mach number enhance the growth of
instabilities in the shear layer surrounding the tail
(Fig.~\ref{fig:various}d). A model with a lower cloud density contrast
($\chi=125$ instead of $10^{3}$) still produces a tail
(Figs.~\ref{fig:various}e-g). Tails still form behind clouds with
smoother density profiles and when the shell is curved
(not shown).

\section{Discussion}
The crushing of clouds by isothermal shells has been investigated only
a few times in the literature
\citep{Tenorio-Tagle:1984b,Rozyczka:1987,Leao:2009}. While these works
have demonstrated that a tail composed of shell material can form in
an interaction with a cloud, the shell, and consequently the focussed
tail, is always much thinner than the cloud radius. Furthermore, some
of the tails are short-lived while others are soon dominated by
ablated material. In contrast, we emphasize that the models presented
here have long-lived tails of thickness comparable to the size of the
cloud.


\subsection{Planetary Nebulae}
Cometary tail-like structures, formed behind dense molecular clouds,
are seen in many PNe. The best studied are those of the highly evolved
PNe NGC\,7293 (the Helix Nebula). The clouds are ionized on the parts
of their surfaces exposed to the ionizing flux from the central star,
and the tails point radially away from it. The tails often contain
molecular material \citep[e.g.,]{Matsuura:2009}. Two different models
have been proposed to explain the tails. In ``shadow'' models the tail
forms due to the shielding of the direct ionizing radiation field of
the central star \citep[e.g.,][]{Lopez-Martin:2001,ODell:2005}. In
contrast, ``stream-source'' models assume that the tail forms from
material photoablated from the cloud
\citep*{Dyson:1993,Falle:2002,Pittard:2005,Dyson:2006}.

The correct model is still disputed
\citep*[e.g.,][]{Dyson:2006,ODell:2007}.  However, observations of the
dynamics of the tails favour stream-source models: i) there is no
evidence for significant ionized gas velocities perpendicular to the
tails \citep{Meaburn:1998}, in contrast to the shadow model of
\citet{Lopez-Martin:2001}; ii) the flow accelerates along the tails
\citep[by about $8-14\,\kmps$, see][]{Meaburn:2010}. While our models
are not specifically of the Helix tails, the velocity increase along
the tail is similar to the observations, which are suggestive of a
linear velocity gradient (Meaburn, private communication).

\subsection{Superbubbles, Starbursts and Superwinds}
The interaction of flows with clouds may be a key mechanism for
producing the broad emission wings to H$\alpha$ line profiles seen in
regions containing concentrations of massive stars including super
star clusters and many giant H{\sc II} regions, such as 30 Doradus
\citep*{Chu:1994,Melnick:1999}. \citet{Westmoquette:2007a,Westmoquette:2007b}
concluded that the broad component arises in a turbulent boundary
layer at the interface between hot gas flowing past cold gas stripped
from clouds. However, subsequent modelling has indicated that
unrealistically high flow speeds are needed
\citep{Binette:2009}. Instead, we believe that the broad emission
wings reflect the acceleration of material along the tail and not just
turbulent motions within the mixing layer.  

Beautiful filamentary structures are also seen in starburst superwinds
\citep[e.g.,][]{Cecil:2001,Ohyama:2002}. While these may be material
stripped from denser clouds, as seen in simulations
\citep[e.g.,][]{Cooper:2008}, the results in this paper
indicate that some of the filaments may in fact have been formed
directly out of an overrunning shell. 

\section{Conclusions}
We have demonstrated a new mechanism for the formation of tails behind
dense clouds which involves the removal and trailing of material from
an overrunning isothermal shell. The mechanism appears robust to a range of
shell thicknesses and Mach numbers, and the cloud density contrast,
though these parameters influence the tail's properties.

\label{lastpage}

\end{document}